\documentstyle[11pt,paspconf,epsf]{article}

\newcommand{\AmS}{{\protect\the\textfont2
  A\kern-.1667em\lower.5ex\hbox{M}\kern-.125emS}}

\def\simlt{\hbox{ \rlap{\raise 0.425ex\hbox{$<$}}\lower 0.65ex\hbox{$\sim$} }}
\def\simgt{\hbox{ \rlap{\raise 0.425ex\hbox{$>$}}\lower 0.65ex\hbox{$\sim$} }}
\def\that{{\hat t}}
\def\umin{u_{\rm min}}

\def\msun{ {\rm M_\odot} }

\def\apj{{\rm ApJ}}
\def\apjl{{\rm ApJL}}

\def\aap{{\rm A\&A}}

\def\etal{{\it et al.}}

\def\pac{Paczy{\'n}ski}

\begin{document}

\title{Planetary Microlensing from the MACHO Project}

\author{D.P.Bennett\rlap,\altaffilmark{1,2,3}
C.Alcock\rlap,\altaffilmark{2,3}
R.A.Allsman\rlap,\altaffilmark{4}
D.Alves\rlap,\altaffilmark{3,5}
T.S.Axelrod\rlap,\altaffilmark{3,4}
A.Becker\rlap,\altaffilmark{2,6}
K.H.Cook\rlap,\altaffilmark{3}
K.C.Freeman\rlap,\altaffilmark{4}
K.Griest\rlap,\altaffilmark{2,7}
M.J.Lehner\rlap,\altaffilmark{2,7}
S.L.Marshall\rlap,\altaffilmark{3}
D.Minniti\rlap,\altaffilmark{3}
B.A.Peterson\rlap,\altaffilmark{4}
M.R.Pratt\rlap,\altaffilmark{2,6}
P.J.Quinn\rlap,\altaffilmark{8}
S.H.Rhie\rlap,\altaffilmark{1,2,3}
A.W.Rodgers\rlap,\altaffilmark{4}
C.W.Stubbs\rlap,\altaffilmark{2,6}
W.Sutherland\rlap,\altaffilmark{9}
T.Vandehei\rlap,\altaffilmark{2,7} and
D.Welch\altaffilmark{10}
}


\altaffiltext{1}{Department of Physics, University of Notre Dame, 
              Notre Dame, IN 46556}
\altaffiltext{2}{Center for Particle Astrophysics, University of California,
        Berkeley, CA 94720}
\altaffiltext{3}{Lawrence Livermore National Laboratory, Livermore, CA 94550}
\altaffiltext{4}{Mount Stromlo and Siding Springs Obs.,
    Australian Nat.~Univ., Weston, ACT 2611, Australia}
\altaffiltext{5}{Department of Physics, University of California, Davis, 
                 CA 95616}
\altaffiltext{6}{Departments of Physics and Astronomy, University of
    Washington, Seattle, WA 98195}
\altaffiltext{7}{Department of Physics, University of
        California San Diego, La Jolla, CA 92093-0350}
\altaffiltext{8}{European Southern Observatory, Garching, Germany}
\altaffiltext{9}{Department of Physics, University of Oxford,
    Oxford OX1 3RH, U.K.}
\altaffiltext{10}{Departments of Physics and Astronomy, McMaster Univ.,
    Hamilton, Ont., Canada L8S 4M1}



\begin{abstract}
We present the lightcurves of two microlensing events from the MACHO
Project data that are likely to be due to lenses with masses similar
to Jupiter's mass. Although the MACHO Project survey data are not
sufficient to definitively establish the identification of planetary
mass lenses in these cases, observations by microlensing follow-up networks
such as GMAN and PLANET should be able to definitively determine the
planetary nature of similar events which may occur in the near future.
\end{abstract}




\section{Introduction}

Gravitational Microlensing has been demonstrated to be a powerful 
observational tool to study populations of stellar or planetary mass
objects which emit little detectable radiation. To date the major
emphasis of gravitational microlensing survey teams has been the
determination of the composition of the Galactic dark matter 
(Alcock, \etal 1996a, 1997b, Ansari, \etal 1996,), but microlensing 
observations toward the Galactic bulge have yielded a surprisingly 
high microlensing rate (Alcock, \etal 1997a, Bennett, \etal 1995,
Udalski, \etal 1994a). This has important implications for the structure
of the Galaxy, but it also yields a relatively large sample of
microlensing events that can be used for other studies.

One of the most exciting possibilities is to use microlensing as a tool to
search for planets orbiting the lensing stars (Mao \& \pac 1991, Gould \&
Loeb 1992). Microlensing is unique among ground based planetary search
techniques for its ability to detect low mass planets (Elachi
1995; Bennett \& Rhie 1996); its sensitivity extends down to an Earth mass.
The microlensing lightcurve deviations caused by planets are generally
quite brief and they will affect only a fraction of all microlensing
events even if every lens star hosts its own planetary system. For these
reasons, microlensing planet searches require real-time event detection
(Alcock, \etal 1996b, Udalski, \etal 1994b)
and frequent microlensing event follow-up observations in order
to have high sensitivity to planetary lensing events. It
is still possible to detect planetary signals with microlensing survey
observations, and in this paper we present two events from the MACHO
Project Galactic bulge data which are likely to have been caused by
lenses with masses close to $M_{\rm Jup}$ (Jupiter's mass).

\section{Events}

\begin{figure}
\plottwo{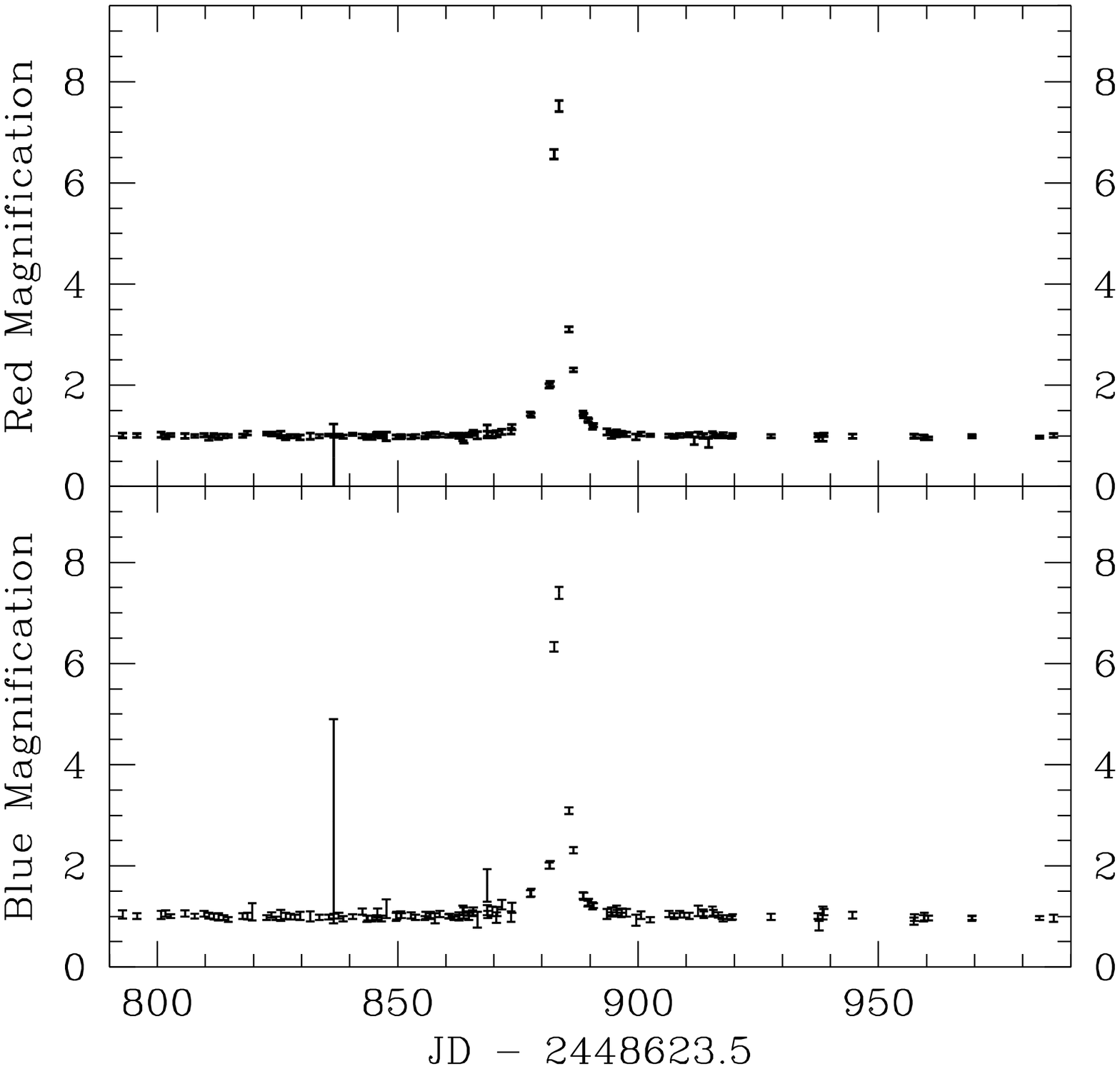}{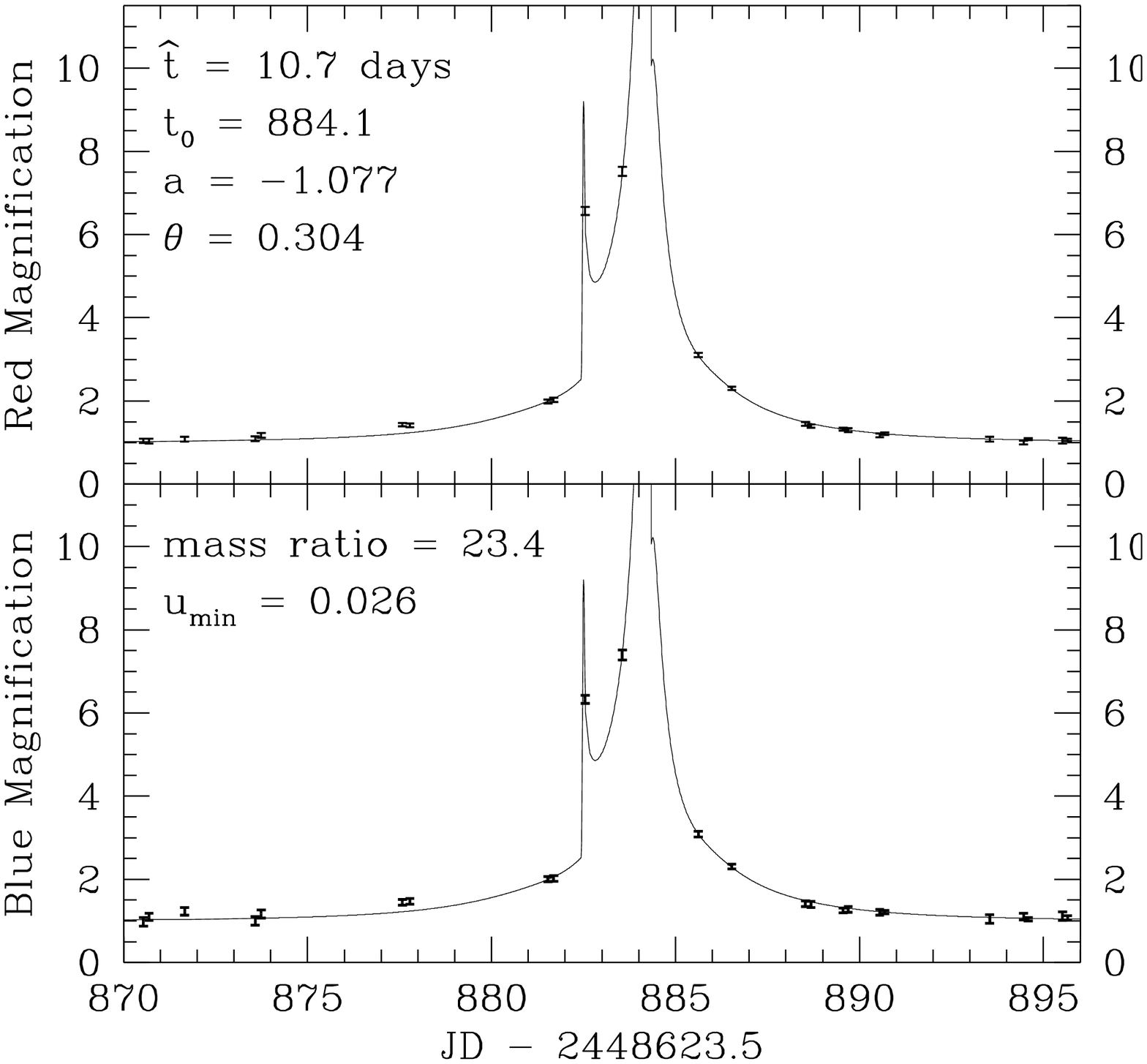}
\caption{The dual-color lightcurve of event 94-BLG-4 during the 1994 
Galactic bulge season and a close-up of the lightcurve showing the binary 
lens fit.\label{fig-a} }
\end{figure}

Figure 1 shows the lightcurve of event 94-BLG-4 from the 1994 bulge season.
This star is a clump giant star with $R = 16.7$ and $V-R = 1.1$ which has
maintained constant brightness during the 1993 through 1996 bulge seasons
with the exception of the short period of brightening shown in Figure 1.
This lightcurve shows a unique brightening which is achromatic but
also asymmetric, and it is also well explained by the binary microlensing
fit shown. The parameters for the binary fit are shown. $\that$ is the
Einstein diameter crossing time; $t_0$ is the time of closest alignment
between the source and lens center of mass; $a$ is the separation 
of the lens components in units of the Einstein radius; $\theta$ 
is the angle between
the motion of the source relative to the line separating the lenses; and
$\umin$ is the transverse distance of closest approach between the source and
lens center of mass. The $\chi^2=430.8$ for the fit shown with 648 data
points and 8 parameters. If we add two other parameters to allow for 
a blended source, the $\chi^2$ is not improved significantly, and we find
that the amount of unlensed light superimposed upon the lensed source
is limited to less than $3\%$ (as expected for a clump giant source).

For comparison, $\chi^2=2835$ for a 5 parameter single lens fit, and if
we arbitrarily remove both the blue and red measurements at day 882.5,
we obtain $\chi^2=491.9$ for the single lens fit. Thus, while we have
undersampled the deviation from the best single lens fit, the deviation is
not completely confined to the pair of data points obtained in the observation
at day 882.5. Formally, the binary fit constrains both the event timescale
and the lens mass ratio quite accurately-to better than $3\%$. The value
$\that = 10.7\,$days indicates a lens mass of $0.04\,\msun$ with a $1-\sigma$
uncertainty of a factor of 3, but the overall $\that$ distribution suggests
that the mass is not much less than $0.1\msun$. This indicates that the
mass of the secondary lens is likely to have $m\sim 5M_{\rm Jup}$ with
a factor of 3 uncertainty.
Thus, the most likely explanation of this event is that the
lens is an M-dwarf system with
a giant planet at a projected separation of (very) roughly 1AU.

\begin{figure}
\plottwo{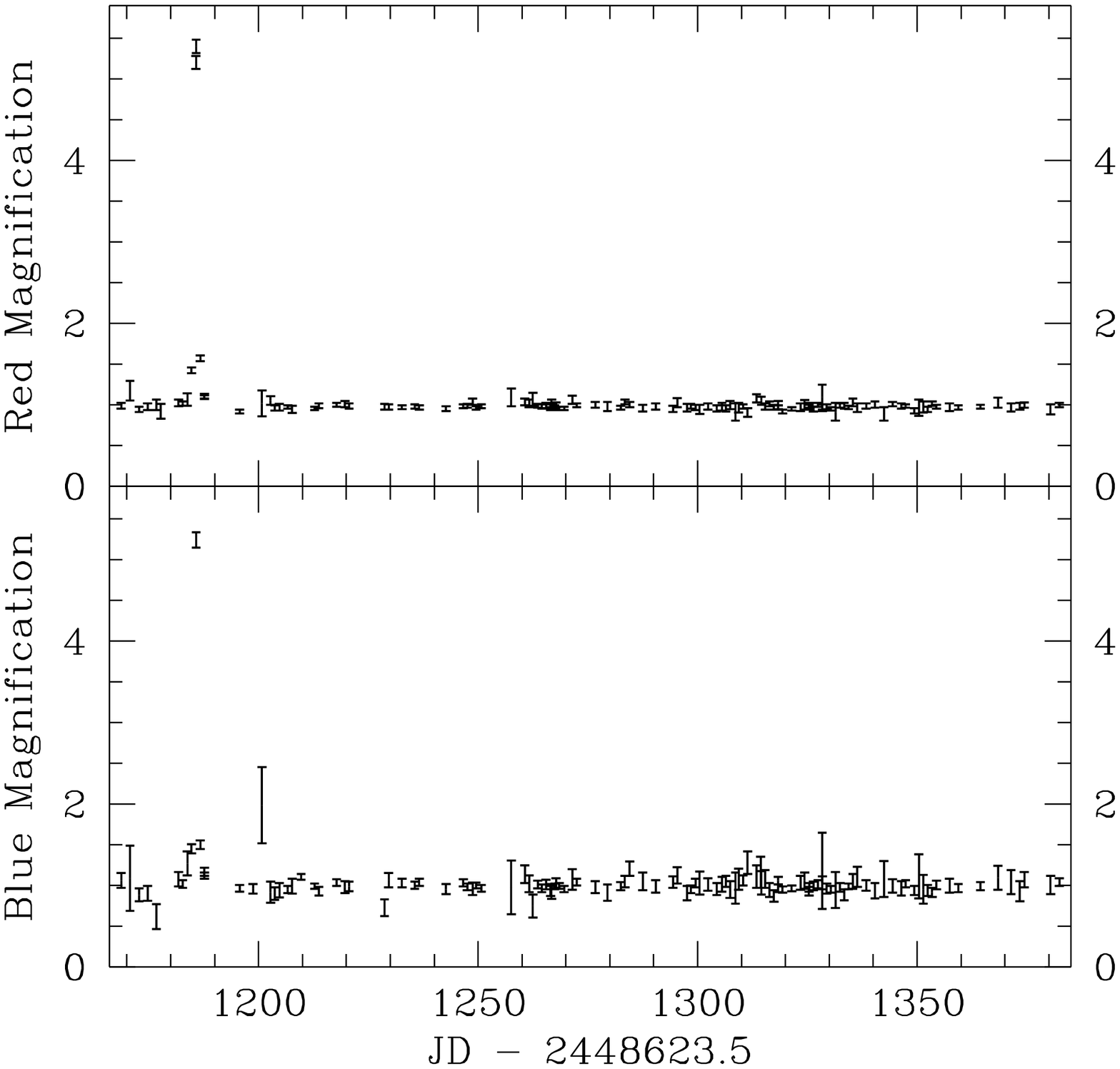}{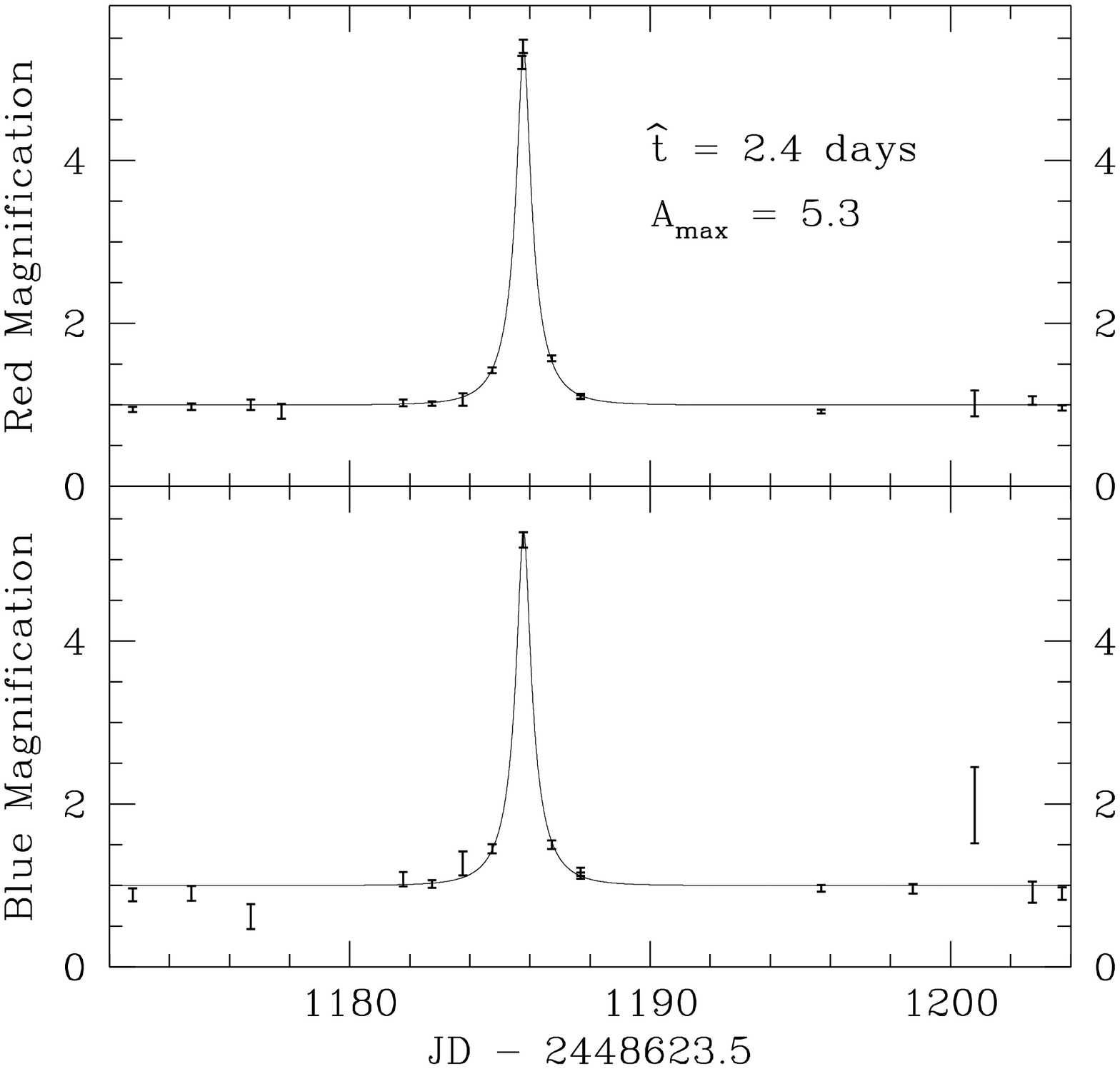}
\caption{The dual-color lightcurve of event 95-BLG-3 during the 1995
Galactic bulge season and a close-up of the lightcurve showing the
lens fit.\label{fig-b} }
\end{figure}

Figure 2 shows the lightcurve of the shortest event ever seen by the MACHO
collaboration with $\that=2.4\,$days. This event was detected with our
alert system, but it also passes the cuts used in our analysis of the '93
bulge data set. If we apply the standard relationship
between $\that$ and lens mass we find a most likely lens mass of about
$2M_{\rm Jup}$, but perhaps we should not use the ``most likely"
mass formula for our shortest event. Couldn't this event be part of
the tail of the event timescale distribution caused by more massive
lenses? The timescale distribution of events from two bulge seasons is
shown in Figure 3. If we assume that mass distribution of lenses has
a lower cutoff (of order $0.1\msun$), then it follows that
the distribution of detected events will scale as $\that^3$ for small
$\that$. (We have used the fact that our event detection efficiency
scales as $\that$ for small $\that$.) This implies that the number of
events with $\that < \that_c$ should scale as $\that_c^4$. 
We can now compare this prediction to the timescale distribution
shown in Figure 3. The '93 data set has 1 event with $\that < 10\,$days
and 12 events with $\that < 20\,$days while the '93$+$'95 data set has
5 events with $\that < 10\,$days and 24 events with $\that < 20\,$days.
Scaling from these numbers with the $\that_c^4$ scaling law implies that 
we should expect to detect between 0.003 and 0.01 events per year with
$\that < 2.5\,$days, so it is unlikely for us to have detected such an
event as a part of the short timescale tail of stellar mass lenses.
Thus, we can treat this event as a part of a separate population and
the mass estimate of $2M_{\rm Jup}$ is a reasonable one. If it is
a planet, then it would have to either be in a distant orbit with a
projected separation of $>5$ or $10\,$AU, or it could be a planet that
has been removed from the planetary system it was born in.

\begin{figure}
\plotfiddle{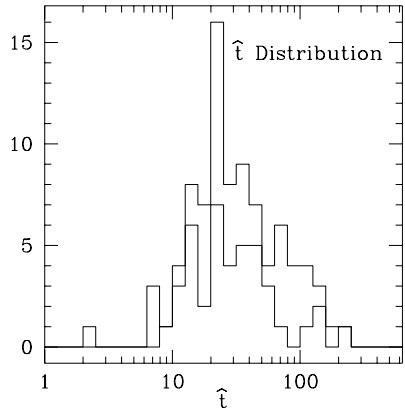}{2.0truein}{0}{28}{28}{-100}{-50}
\caption{The $\that$ distributions of the 1993 MACHO bulge events and
the combined 1993 and 1995 bulge events (bold) are shown.}
\end{figure}

\section{Conclusions}

Unfortunately, we cannot treat either of these two events as definitive
detections of planetary mass objects. For the first event, 94-BLG-4, 
we would require additional observations to fully characterize the
binary lightcurve and to definitively establish that our fit is the
correct one. For event 95-BLG-3, additional observations taken less than
24 hours after the event peak could have determined if the finite size of
the source star was resolved which would have established this event
as a {\it bona fide} planetary mass lensing event. Had this event occurred
in 1996, we would have recognized this event in time to request
follow-up observations, but in early 1995
our alert system was operating with a time lag of about 30 hours.
At present, the time lag for MACHO alerts is typically less than 6 hours.
When similar events occur in the future, we can look forward to prompt
alert announcements and to higher quality data sets from the ever 
expanding microlensing follow-up teams such as GMAN and PLANET
(Pratt, \etal 1995, Albrow, \etal 1995)

\acknowledgments
Work performed at LLNL is supported by the DOE under contract W7405-ENG-48.
Work performed by the CfPA personnel is supported in part by the 
Office of Science and Technology Centers of
NSF under cooperative agreement AST-8809616.
Work performed at MSSSO is supported by the Bilateral Science
and Technology Program of the Australian Department of Industry, Technology
and Regional Development.


%
%

%

\end{document}